\documentclass[reprint, amsmath, amssymb, aps, floatfix, prl, superscriptaddress]{revtex4-2}

\usepackage{graphicx}
\usepackage{dcolumn}
\usepackage{bm}
\usepackage{hyperref}
\usepackage{appendix}

\begin{document}

\preprint{APS/123-QED}

\title{Hydrodynamic and Electromagnetic Discrepancies between Neutron Star and Black Hole Spacetimes}

\author{Jonathan Gorard}
\email{gorard@princeton.edu}
\affiliation{Princeton University, Princeton, NJ 08544, USA}
\affiliation{Princeton Plasma Physics Laboratory, Princeton, NJ, 08540, USA}

\author{James Juno}
\affiliation{Princeton Plasma Physics Laboratory, Princeton, NJ, 08540, USA}

\author{Ammar Hakim}
\affiliation{Princeton Plasma Physics Laboratory, Princeton, NJ, 08540, USA}

\date{\today}

\begin{abstract}
The exterior spacetime geometry surrounding an uncharged, spinning black hole in general relativity depends only upon its mass and spin. However, the exterior geometry surrounding any other rotating compact object, for example a neutron star, will generally depend upon higher moments in its multipole expansion, which will in turn be dependent upon the object's equation of state. Using general relativistic hydrodynamics and electrodynamics simulations, we illustrate that the presence or absence of these higher moments (assuming a physically realistic neutron star equation of state) has a significant qualitative effect near the surface of the compact object on the dynamics of unmagnetized accretion, and a smaller quantitative effect on the electromagnetic field configuration of its magnetosphere. In some places, the discrepancies in energy-momentum density are found to reach or exceed 50\%, with electric field strength discrepancies in excess of 10\%. We argue that many of these differences are likely to be amplified by the inclusion of more sophisticated plasma physics models, and are therefore likely to be relevant for the dynamics of gravitational collapse, and potentially also for particle acceleration and jet launching. These discrepancies suggest important limitations regarding the use of the Kerr metric when performing numerical simulations around neutron stars.
\end{abstract}

\maketitle

Birkhoff's theorem\cite{birkhoff_relativity_1927} implies that the exterior spacetime geometry surrounding any spherically-symmetric, non-rotating object in general relativity is described by the Schwarzschild metric. Therefore, general relativistic simulations around non-rotating, spherically-symmetric compact objects may safely be performed within a Schwarzschild background, irrespective of whether the compact object itself is a black hole, neutron star, white dwarf, quark star, or other exotic object. However, in the case of rotating astrophysical bodies, Birkhoff's theorem no longer applies, and so one finds that the exterior spacetime geometry surrounding a generic, rigidly-rotating, axisymmetric object (as described, for instance, by the Hartle-Thorne metric\cite{hartle_slowly_1967}\cite{hartle_slowly_1968} in the slowly-rotating limit ${\Omega^2 \ll \frac{G M}{R^3}}$) is not necessarily equivalent to the exterior spacetime geometry surrounding a spinning black hole (as described by the Kerr metric). More concretely, the no-hair theorem\cite{heusler_black_2003} implies that the spacetime surrounding an uncharged black hole is fully characterized by the first two moments of its multipole expansion, namely its mass $M$ and angular momentum $J$, while the spacetime surrounding a more general uncharged astrophysical body may also depend upon higher moments, including its mass quadrupole ${M_2}$, spin octupole ${S_3}$, and mass hexadecapole ${M_4}$\cite{chandrasekhar_slowly_1974}\cite{bradley_quadrupole_2009}\cite{paschalidis_rotating_2017}.

Nevertheless, many state-of-the-art general relativistic simulations of accretion, jet production, gravitational collapse, and magnetospheric interactions surrounding rotating neutron stars use the Schwarzschild or Kerr metrics to represent their underlying spacetime geometry\cite{cikintoglu_grmhd_2022}\cite{aps2021}\cite{parfrey_accreting_2024}, largely for reasons of simplified numerical implementation: such metrics admit a straightforward description in terms of the \textit{horizon-penetrating} Kerr-Schild coordinate system\cite{font_horizon-adapted_1998}, which prevents numerical instabilities and unphysical divergences from appearing near event horizons. At large distances from the compact object, this is clearly a reasonable approximation, since the Kerr metric is known to agree with more general axisymmetric metrics, such as the Hartle-Thorne metric, in the weak-field limit. However, we demonstrate within this Letter that there exist large qualitative discrepancies in the behaviors of unmagnetized fluids close to the compact object, as well as smaller quantitative discrepancies in the behaviors of electromagnetic fields, depending upon whether one's simulation framework uses a Kerr metric or a physically realistic neutron star metric. Near the approximate surface of the compact object, we find that the discrepancies in fluid momentum densities can reach 30--40\%, those in fluid energy densities can reach 50--70\%, and those in electric field strengths can reach 10--12\%, when comparing physically realistic neutron star metrics against black hole metrics of equivalent mass and spin. We perform these general relativistic hydrodynamics and electrodynamics simulations assuming a typical neutron star angular momentum of ${J = 0.4 M^2}$, and typical neutron star equation of state parameters of moderate to low stiffness (corresponding to quadrupolar deformability values of ${\alpha = 5}$ and ${\alpha = 8}$, respectively). Simulations are performed using the tetrad-based general relativistic solvers\cite{gorard2025} implemented within the \textsc{Gkeyll} simulation framework. Einstein summation convention, ${\left( -, +, +, + \right)}$ metric signature, and geometrized units with ${G = c = 1}$, are assumed throughout. Greek indices are used to index the spacetime coordinates ${\left\lbrace x^0, x^1, x^2, x^3 \right\rbrace}$, while Latin indices are used to index the spatial coordinates ${\left\lbrace x^1, x^2, x^3 \right\rbrace}$.

We opt to use the approximate neutron star metric of Pappas\cite{pappas_accurate_2017}, since this metric is known to agree with numerical solutions to the Einstein field equations for physically realistic neutron stars with very high accuracy within the particular regions of parameter space that we will be simulating (i.e. quadrupolar deformability parameters ${\alpha \leq 8}$, angular momentum parameters ${J \leq 0.5 M^2}$). Following Pappas, we use the cylindrical Weyl-Lewis-Papapetrou coordinates ${\left( t, \rho, \varphi, z \right)}$ to construct a general axisymmetric spacetime line element of the form:

\begin{align*}\nonumber
d s^2 &= g_{\mu \nu} \, d x^{\mu} \, d x^{\nu}\\
&= -f \left( d t - \omega \, d \varphi \right)^2 + f^{-1} \left[ e^{2 \gamma} \left( d \rho^2 + d z^2 \right) + \rho^2 \, d \varphi^2 \right],
\end{align*}
with metric functions ${f \left( \rho, z \right)}$, ${\omega \left( \rho, z \right)}$, and ${\gamma \left( \rho, z \right)}$ as defined in the \hyperref[sec:pappas_metric]{End Matter}. To facilitate our numerical simulations, we decompose the spacetime metric into its ADM/${3 + 1}$ form\cite{arnowitt_dynamical_1959}:

\begin{align*}\nonumber
d s^2 &= g_{\mu \nu} \, d x^{\mu} \, d x^{\nu}\\
&= \left( - N^2 + N_i N^i \right) d t^2 + 2 N_i \, d t \, d x^i + \gamma_{i j} \, d x^i \, d x^j
\end{align*}
with the lapse function $N$ and the angular component of the shift vector ${N^{\varphi}}$ given by:

\begin{equation}\nonumber
N = \frac{\sqrt{f} \rho}{\sqrt{\rho^2 - f^2 \omega^2}}, \qquad N^{\varphi} = - \frac{f^2 \omega}{f^2 \omega^2 - \rho^2},
\end{equation}
respectively, with ${N^{\rho} = N^z = 0}$, and with the spatial line element given by:

\begin{equation*}\nonumber
d l^2 = \gamma_{i j} \, d x^i \, d x^j = f^{-1} e^{2 \gamma} \left( d \rho^2 + d z^2 \right) + \rho^2 \omega^2 \, d \varphi^2.
\end{equation*}
Henceforth, we shall refer to any observer whose four-velocity ${\mathbf{u}}$ is normal to the spacelike hypersurfaces described by ${\gamma_{i j}}$ as an \textit{Eulerian observer}.

Following Pappas and Apostolatos\cite{pappas_effectively_2014}, we parameterize the higher multipole moments ${M_2}$, ${S_3}$, and ${M_4}$ in terms of deviations from the Kerr metric:

\begin{equation}\nonumber
M_2 = - \alpha j^2 M^3, \qquad S_3 = - \beta j^3 M^4, \qquad M_4 = \gamma j^4 M^5,
\end{equation}
where ${j = {J}/{M^2}}$ is the dimensionless spin parameter, and the degenerate case ${\alpha = \beta = \gamma = 1}$ corresponds to the Kerr metric. For physically realistic neutron star equations of state, the higher moments are not algebraically independent, but rather can be expressed in terms of the coefficient ${\alpha}$ in front of the mass quadrupole (i.e. the quadrupolar deformability) as\cite{pappas_effectively_2014}\cite{yagi_effective_2014}:

\begin{equation}\nonumber
\beta = \left( - 0.36 + 1.48 \left( \sqrt{\alpha} \right)^{0.65} \right)^3,
\end{equation}
\begin{equation}\nonumber
\gamma = \left( - 4.749 + 0.27613 \left( \sqrt{\alpha} \right)^{1.5146} + 5.5168 \left( \sqrt{\alpha} \right)^{0.22229} \right)^4.
\end{equation}
We select quadrupolar deformability values of ${\alpha = 5}$ and ${\alpha = 8}$ for our simulations, with the former corresponding to a typical value for a moderately stiff equation of state for the nuclear matter (such as the FPS equation of state, with ${\alpha = 4.209}$\cite{cook_rapidly_1994}), and the latter corresponding to a typical value for a lower-mass model. We also select a dimensionless spin of ${j = 0.4}$, corresponding to the estimated spin of PSR J1748-2446ad\cite{hessels_radio_2006} (a pulsar with a measured frequency of 716 Hz), since this value is both astrophysically plausible and lies comfortably below the typical breakup speed ${j \sim 0.7}$ for standard neutron star equations of state\cite{wang_targeted_2024}. We place the excision boundary for our numerical simulations at the Schwarzschild radius ${r = 2M}$ to prevent prohibitive time-step restrictions due to divergences in the metric and its derivatives. However, we note that the physical surface of a neutron star above ${\sim}$one solar mass is likely to lie somewhere between ${r = 3M}$ and ${r = 4M}$\cite{haensel_apparent_2001}. For the purposes of our numerical comparisons, we shall assume that the neutron star surface lies at ${r = 3M}$ exactly, and shall consider all behavior in the ``interior'' region ${2M \leq r \leq 3M}$ to be unphysical. Black hole spacetimes (for comparison purposes) are simulated by setting ${\alpha = \beta = \gamma = 1}$ within the Pappas metric.

\begin{figure*}[htp]
\includegraphics[width=0.45\textwidth]{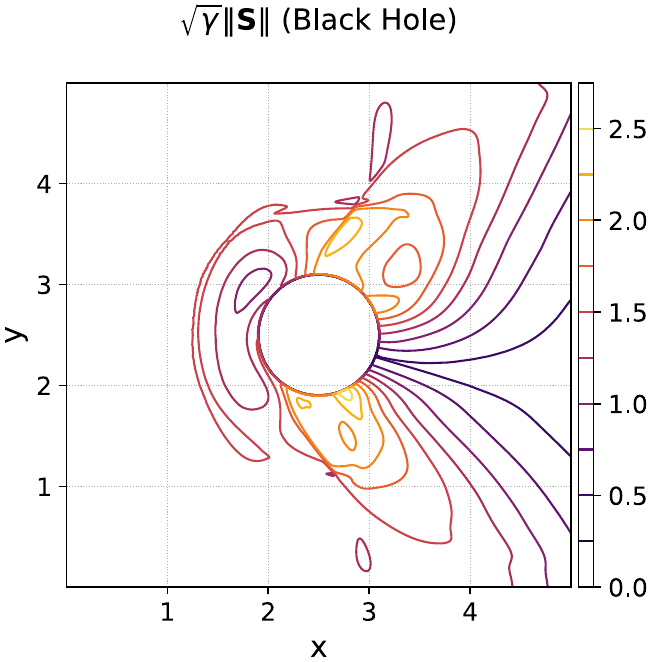}
\includegraphics[width=0.45\textwidth]{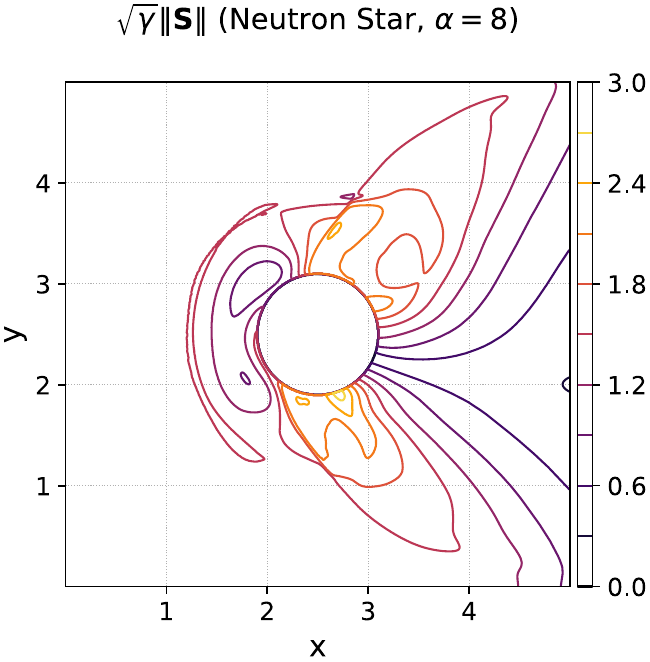}
\caption{The configuration of momentum density contours (as perceived by an Eulerian observer) at time ${t = 15}$ for the ideal gas accretion problem onto a compact object with ${v_{\infty} = 0.3}$ and ${J = 0.4 M^2}$, assuming a black hole metric (left) and a neutron star metric with ${\alpha = 0.8}$ (right).}
\label{fig:BHLMomentumContours}
\end{figure*}

\begin{figure*}[htp]
\includegraphics[width=0.45\textwidth]{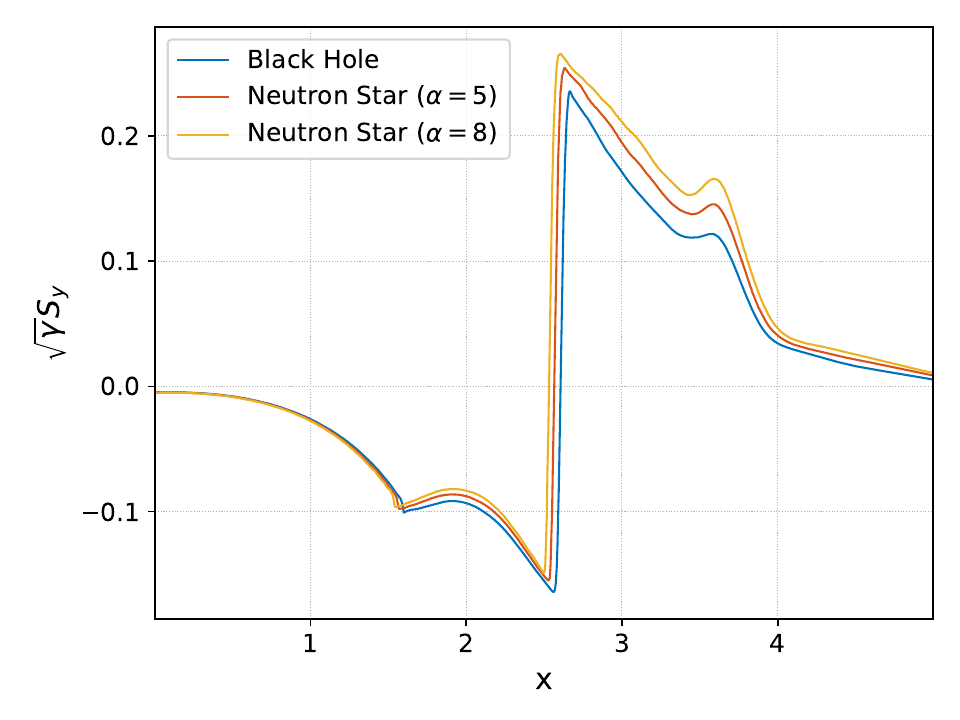}
\includegraphics[width=0.45\textwidth]{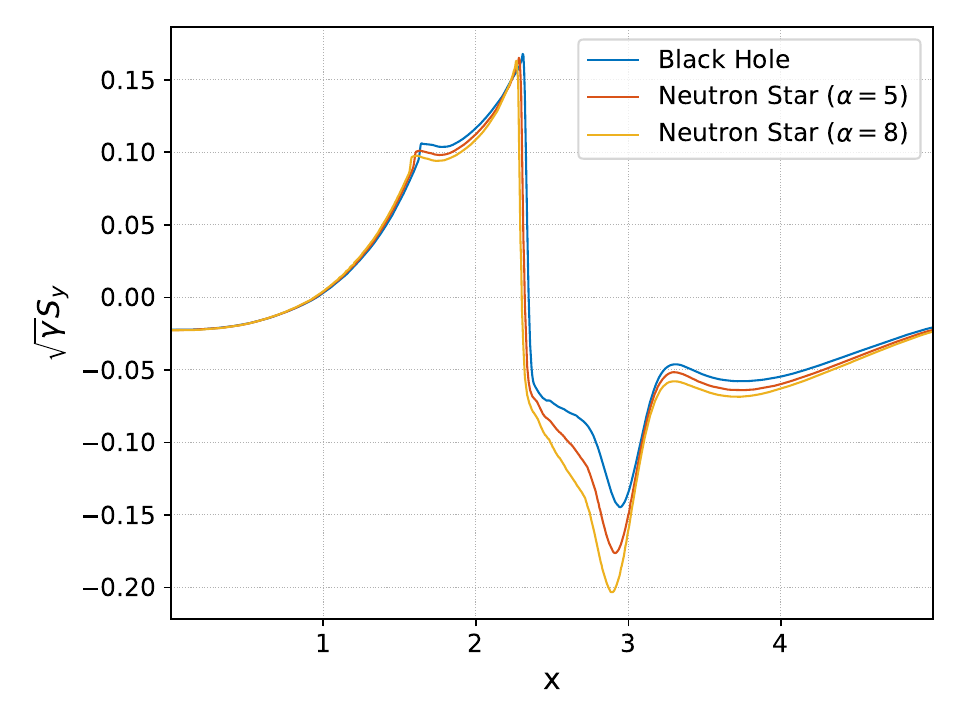}
\caption{Cross-sectional profiles of the $y$-component of the fluid momentum density ${\mathbf{S}}$ (as perceived by an Eulerian observer) at time ${t = 15}$ for the ideal gas accretion problem onto a compact object with ${v_{\infty} = 0.3}$ and ${J = 0.4 M^2}$. Cross-sections taken at ${y = 3.4}$ (left) and ${y = 1.6}$ (right), representing the co-rotating and counter-rotating surfaces of the compact object, respectively, assuming a neutron star radius of ${r \sim 3M}$.}
\label{fig:BHLMomentumCrossSection}
\end{figure*}

First, we simulate wind accretion of an ideal gas with adiabatic index ${\Gamma = \frac{5}{3}}$ (i.e. a monatomic gas) onto a rotating compact object of mass ${M = 0.3}$ and angular momentum ${J = 0.4 M^2}$ centered at ${\left( x, y \right) = \left( 2.5, 2.5 \right)}$, assuming a square domain ${\left( x, y \right) \in \left[ 0, 5 \right] \times \left[ 0, 5 \right]}$. The setup of this problem is based on that of Font, Ib\'a\~nez and Papadopoulos\cite{font_non-axisymmetric_1999}, and was previously simulated using \textsc{Gkeyll} in \cite{gorard2025}. We assume the fluid density, pressure, and velocity at spatial infinity to be given by ${\rho_{\infty} = 3}$, ${p_{\infty} = 0.05}$ and ${v_{\infty} = 0.3}$, respectively, while in the remainder of the domain one has ${\rho_0 = p_0 = 0.01}$ and ${v_0 = 0}$. A summary of the equations being solved, as well as the notation used, can be found in the \hyperref[sec:grhd_equations]{End Matter}. The simulation is run using a spatial discretization of ${1024 \times 1024}$ cells, a CFL coefficient of 0.95, and up to a final time of ${t = 15}$. Figure \ref{fig:BHLMomentumContours} shows the qualitative difference in the fluid momentum density contours ${\sqrt{\gamma} \left\lVert \mathbf{S} \right\rVert}$ (as perceived by an Eulerian observer) when using a black hole metric and using a neutron star metric with ${\alpha = 8}$. Figure \ref{fig:BHLMomentumCrossSection} shows the quantitative discrepancy in the $y$-component of the fluid momentum density ${\sqrt{\gamma} S_y}$ between black hole and neutron star metrics with ${\alpha = 5}$ and ${\alpha = 8}$, via cross-sectional profiles through the ${y = 3.4}$ and ${y = 1.6}$ axes. Assuming a neutron star radius of ${r \sim 3M}$, these correspond to the cross-sectional profiles at the neutron star surface on the co-rotating and counter-rotating sides of the accretion flow, respectively. We see that the peak value of ${\left\lvert \sqrt{\gamma} S_y \right\rvert}$ is modified by around 15--20\% on the co-rotating side, and by around 30--40\% on the counter-rotating side, between the black hole and neutron star cases (with the larger discrepancies occurring for ${\alpha = 8}$). In some regions, we find that the peak value of the relativistic energy density ${\sqrt{\gamma} \tau}$ near the neutron star surface may consequently be modified by up to 50--70\% by the use of a physically realistic neutron star metric over a black hole metric.

\begin{figure*}[htp]
\includegraphics[width=0.45\textwidth]{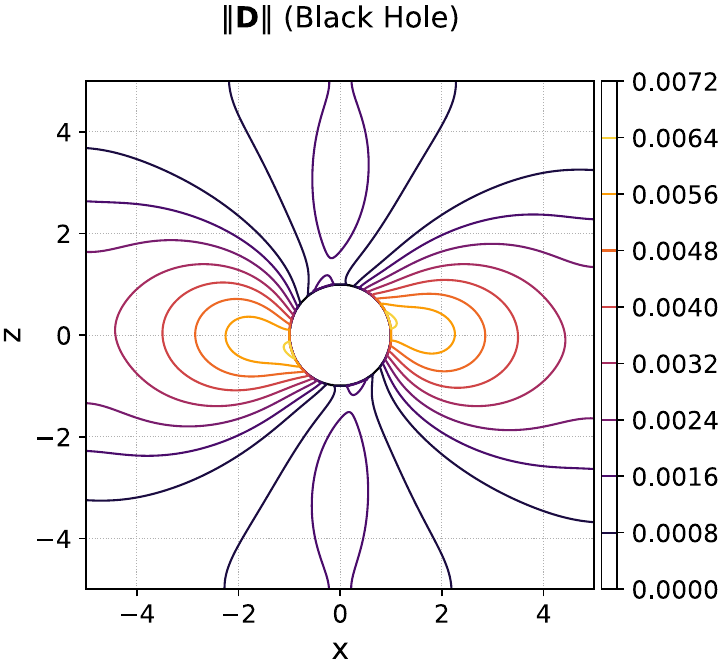}
\includegraphics[width=0.45\textwidth]{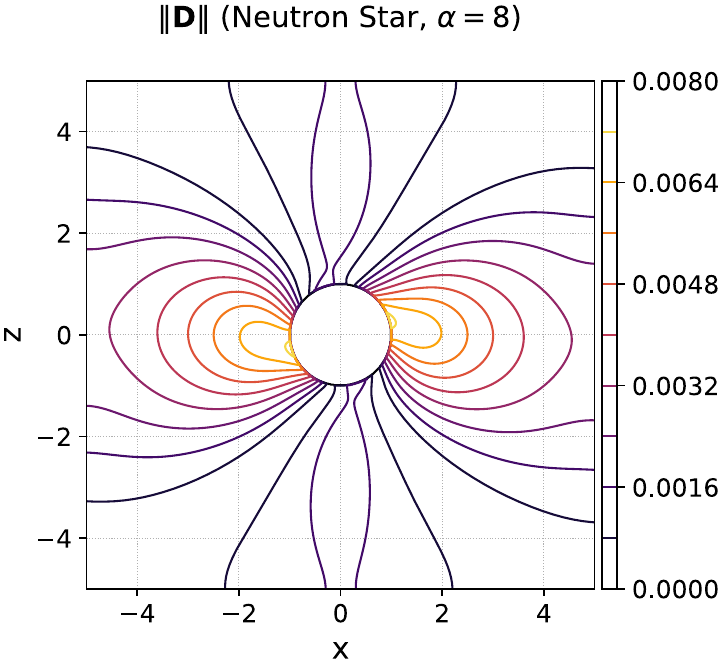}
\caption{The configuration of electric flux surfaces (as perceived by an Eulerian observer) at time ${t = 50}$ for the Wald-type magnetosphere problem around a compact object with ${B_0 = 1}$ and ${J = 0.4 M^2}$, assuming a black hole metric (left) and a neutron star metric with ${\alpha = 0.8}$ (right).}
\label{fig:WaldElectricField}
\end{figure*}

\begin{figure*}[htp]
\includegraphics[width=0.45\textwidth]{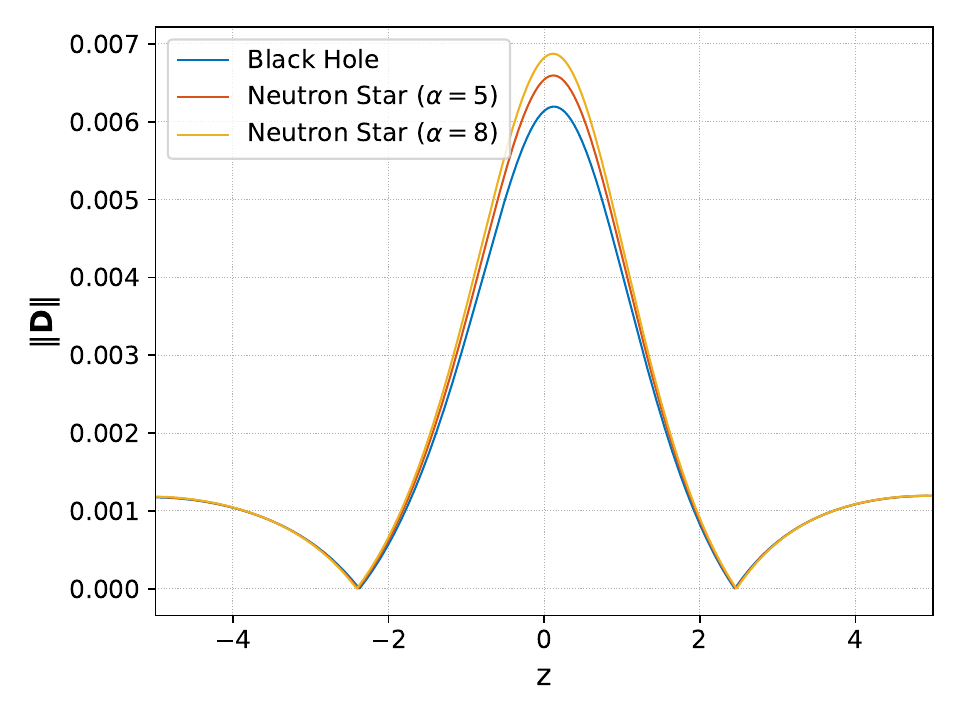}
\includegraphics[width=0.45\textwidth]{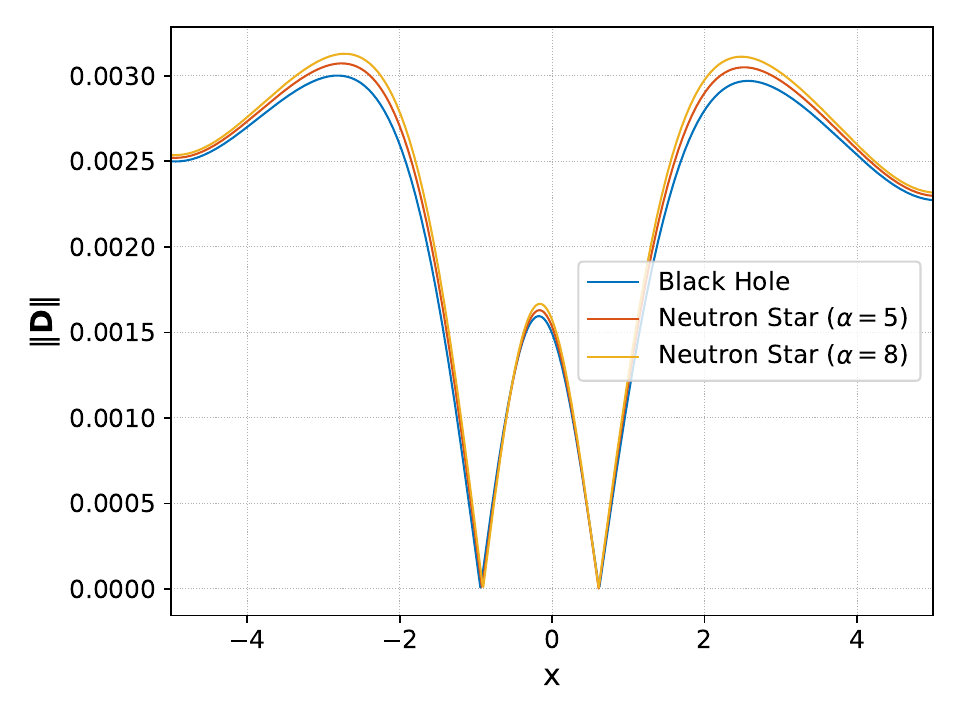}
\caption{Cross-sectional profiles of the electric field strength ${\left\lVert \mathbf{D} \right\rVert}$ (as perceived by an Eulerian observer) at time ${t = 50}$ for the Wald-type magnetosphere problem around a compact object with ${B_0 = 1}$ and ${J = 0.4 M^2}$. Cross-sections taken at ${x = 1.5}$ (left) and ${z = 1.5}$ (right), representing surfaces of the compact object in the axial and equatorial directions, respectively, assuming a neutron star radius of ${r \sim 3M}$.}
\label{fig:WaldElectricCrossSection}
\end{figure*}

Next, we simulate a Wald-type magnetosphere problem\cite{wald_black_1974} for a rotating compact object of mass ${M = 0.5}$ and angular momentum ${J = 0.4 M^2}$, centered at ${\left( x, z \right) = \left( 0, 0 \right)}$, assuming a square domain ${\left( x, z \right) \in \left[ -5, 5 \right] \times \left[ -5, 5 \right]}$. Again, this problem was previously simulated using \textsc{Gkeyll} in \cite{gorard2025}, and again a summary of the equations being solved and the notation being used can be found in the \hyperref[sec:grem_equations]{End Matter}. The object is immersed in an initially-uniform magnetic field of strength ${B_0 = 1}$, aligned with the object's spin axis, and the simulation is again run using a spatial discretization of ${1024 \times 1024}$ cells, a CFL coefficient of 0.95, and now up to a final time of ${t = 50}$. Frame-dragging of the magnetic field lines in the surrounding spacetime causes large parallel electric fields to be induced around the surface of the compact object. Figure \ref{fig:WaldElectricField} shows the qualitative difference in the electric flux surfaces ${\left\lVert \mathbf{D} \right\rVert}$ (as perceived by an Eulerian observer) when using a black hole metric and using a neutron star metric with ${\alpha = 8}$. Figure \ref{fig:WaldElectricCrossSection} shows the quantitative discrepancy in the electric field strength between black hole and neutron star metrics with ${\alpha = 5}$ and ${\alpha = 8}$, via cross-sectional profiles through the ${x = 1.5}$ and ${z = 1.5}$ axes. Assuming a neutron star radius of ${r \sim 3M}$, these correspond to the cross-sectional profiles at the neutron star surface in the axial and equatorial directions, respectively. We find that the peak value of ${\left\lVert \mathbf{D} \right\rVert}$ is modified by around 10--12\% in the axial direction, and around 5--8\% in the equatorial direction, between the black hole and neutron star cases (again, with the larger discrepancies occurring for ${\alpha = 8}$). The relative effects on the magnetic field strength are significantly lower, largely as a consequence of the background magnetic field itself being much stronger.

Although we have shown that the dynamics of unmagnetized accretion are substantially modified by the choice to use a black hole vs. a realistic neutron star metric, the fact that the magnetic field strength near the surface of the compact object does not change by more than a few percent between these two cases means that realistic plasma accretion rates are likely to be much less strongly affected: the magnetic field around a rotating neutron star is typically sufficiently strong that the surrounding plasma is forced to co-rotate with it, and essentially all accretion occurs parallel to the magnetic field\cite{philippov_kramer_2022}. Thus, we can conclude that the Kerr metric remains a reasonable approximation for the spacetime geometry in the context of simulations of quantitative accretion rates around highly magnetized neutron stars, or of electromagnetic field configurations within isolated pulsar magnetospheres. However, for phenomena such as gravitational collapse, which may be highly sensitive to the accretion geometry\cite{sekiguchi_axisymmetric_2005}, these large changes in the energy and momentum densities in the unmagnetized hydrodynamics near the surface of the compact object are likely to be more significant. Moreover, many of these deviations are likely to become further accentuated by buoyancy-driven instabilities such as magnetized Rayleigh-Taylor\cite{parfrey_tchekhovskoy_2024}, kinetic streaming instabilities parallel to the magnetic field\cite{beloborodov_2013}, and other effects resulting from the inclusion of more sophisticated plasma physics models. In effect, even if the plasma is mostly streaming parallel to the magnetic field, the geometry of the plasma’s ultimate deposition onto the compact object will be a function of how the fluid transport is modified by the neutron star metric; the motion of both the co-rotating and streaming plasma close to the compact object, where even small changes to where the matter in-falls can affect the ultimate collapse, will be modified by the higher multipole moments of the exterior spacetime.

Furthermore, the Blandford-Znajek mechanism\cite{blandford_electromagnetic_1977}, which is widely hypothesized to be the underlying mechanism driving the emission of relativistic jets by spinning black holes\cite{parfrey_first-principles_2019}, is also known to have a straightforward analog that applies to the case of jets produced by rotating neutron stars too\cite{parfrey_tchekhovskoy_2017}\cite{das_three-dimensional_2024}. The dynamics of the Blandford-Znajek process (and presumably its neutron star analog) are known to be strongly influenced by the large parallel electric fields ${\left\lvert \mathbf{D} \cdot \mathbf{B} \right\rvert \sim B^2}$ induced by the frame-dragging of magnetic field lines intersecting the compact object. Therefore, we may reasonably expect the 10--12\% discrepancies in the electric field strength near the surface of the object in the axial direction to play at least some role in the dynamics of jet formation and particle acceleration around the object. 

We reiterate that all of our simulations have assumed physically plausible parameters for the neutron star throughout, without extremes of either angular momentum (i.e. ${J \leq 0.4 M^2}$) or equation of state parameter (i.e. ${\alpha \leq 8}$). Hence, it is plausible that, in contrast to simulations of simple accretion and magnetosphere scenarios, simulations of gravitational collapse and particle acceleration close to the surface of the compact object may be markedly affected by the choice to use the Kerr metric over a physically realistic neutron star metric as a description of the underlying spacetime geometry. Thus, not only does the use of this metric give more realistic initial conditions for simulations of collapsing compact objects when coupled to the Einstein field equations, but the physics fidelity of such collapse simulations necessitates the use of an appropriate metric too.

\begin{acknowledgments}
J.G. was partially funded by the Princeton University Research Computing group. J.G., J.J., \& A.H. were partially funded by the U.S. Department of Energy under Contract No. DE-AC02-09CH1146 via an LDRD grant. The development of \textsc{Gkeyll} was partially funded, besides the grants mentioned above, by the NSF-CSSI program, Award Number 2209471.
\end{acknowledgments}

\bibliography{NeutronStarBib}

\begin{appendices}

\onecolumngrid

\section{End Matter}

\subsection{Metric Functions for the Pappas Metric}
\label{sec:pappas_metric}

The particular forms of the metric functions ${f \left( \rho, z \right)}$, ${\omega \left( \rho, z \right)}$, and ${\gamma \left( \rho, z \right)}$ appearing within the general axisymmetric line element:

\begin{equation*}
d s^2 = g_{\mu \nu} \, d x^{\mu} \, d x^{\nu} = - f \left( dt - \Omega \, d \varphi \right)^2 + f^{-1} \left[ e^{2 \gamma} \left( d \rho^2 + d z^2 \right) + \rho^2 \, d \varphi^2 \right],
\end{equation*}
that characterize the Pappas metric\cite{pappas_accurate_2017} are given by:

\begin{multline*}
f \left( \rho, z \right) = 1 - \frac{2 M}{\sqrt{\rho^2 + z^2}} + \frac{2 M^2}{\rho^2 + z^2} + \frac{\left( M_2 - M^3 \right) \rho^2 - 2 \left( M^3 + M_2 \right) z^2}{\left( \rho^2 + z^2 \right)^\frac{5}{2}}\\
+ \frac{2 z^2 \left( - J^2 + M^4 + 2 M_2 M \right) - 2 M M_2 \rho^2}{\left( \rho^2 + z^2 \right)^3} + \frac{A \left( \rho, z \right)}{28 \left( \rho^2 + z^2 \right)^{\frac{9}{2}}} + \frac{B \left( \rho, z \right)}{14 \left( \rho^2 + z^2 \right)^5},
\end{multline*}

\begin{equation*}
\omega \left( \rho, z \right) = - \frac{2 J \rho^2}{\left( \rho^2 + z^2 \right)^{\frac{3}{2}}} - \frac{2 J M \rho^2}{\left( \rho^2 + z^2 \right)^2} + \frac{F \left( \rho, z \right)}{\left( \rho^2 + z^2 \right)^{\frac{7}{2}}} + \frac{H \left( \rho, z \right)}{2 \left( \rho^2 + z^2 \right)^4} + \frac{G \left( \rho, z \right)}{4 \left( \rho^2 + z^2 \right)^{\frac{11}{2}}},
\end{equation*}

\begin{equation*}
\gamma \left( \rho, z \right) = \frac{\rho^2 \left( J^2 \left( \rho^2 - 8 z^2 \right) + M \left( M^3 + 3 M_2 \right) \left( \rho^2 - 4 z^2 \right) \right)}{4 \left( \rho^2 + z^2 \right)^4} - \frac{M^2 \rho^2}{2 \left( \rho^2 + z^2 \right)^2},
\end{equation*}
where:

\begin{multline*}
A \left( \rho, z \right) = \left[ 8 \rho^2 z^2 \left( 24 J^2 M + 17 M^2 M_2 + 21 M_4 \right) + \rho^4 \left( -10 J^2 M + 7 M^5 + 32 M_2 M^2 - 21 M_4 \right) \right.\\
\left. + 8 z^4 \left( 20 J^2 M - 7 M^5 - 22 M_2 M^2 - 7 M_4 \right) \right],
\end{multline*}

\begin{multline*}
B \left( \rho, z \right) = \left[ \rho^4 \left( 10 J^2 M^2 + 10 M_2 M^3 + 21 M_4 M + 7 M_{2}^{2} \right) \right.\\
+ 4 z^4 \left( - 40 J^2 M^2 - 14 J S_3 + 7 M^6 + 30 M_2 M^3 + 14 M_4 M + 7 M_{2}^{2} \right)\\
\left. - 4 \rho^2 z^2 \left( 24 J^2 M^2 - 21 J S_3 + 7 M^6 + 48 M_2 M^3 + 42 M_4 M + 7 M_{2}^{2} \right) \right],
\end{multline*}

\begin{equation*}
H \left( \rho, z \right) = \left[ 4 \rho^2 z^2 \left( J \left( M_2 - 2 M^3 \right) - 3 M S_3 \right) + \rho^4 \left( J M_2 + 3 M S_3 \right) \right].
\end{equation*}

\begin{multline*}
G \left( \rho, z \right) = \left[ \rho^2 \left( J^3 \left( - \left( \rho^4 + 8 z^4 - 12 \rho^2 z^2 \right) \right) + J M \left( \left( M^3 + 2 M_2 \right) \rho^4 - 8 \left( 3 M^3 + 2 M_2 \right) z^4 \right. \right. \right.\\
\left. \left. \left. + 4 \left( M^3 + 10 M_2 \right) \rho^2 z^2 \right) + M^2 S_3 \left( 3 \rho^4 - 40 z^4 + 12 \rho^2 z^2 \right) \right) \right],
\end{multline*}

\begin{align*}\nonumber
F \left( \rho, z \right) = \left[ \rho^4 \left( S_3 - J M^2 \right) - 4 \rho^2 z^2 \left( J M^2 + S_3 \right) \right].
\end{align*}

\subsection{Equations of General Relativistic Hydrodynamics}
\label{sec:grhd_equations}

\textsc{Gkeyll} solves the equations of general relativistic hydrodynamics in hyperbolic, conservation law form following the ${3 + 1}$ \textit{Valencia} formalism\cite{banyuls_numerical_1997}, consisting of conservation laws for baryonic number density:

\begin{equation}\nonumber
\frac{1}{N \sqrt{\gamma}} \left( \frac{\partial}{\partial t} \left( \sqrt{\gamma} D \right) + \frac{\partial}{\partial x^i} \left\lbrace N \sqrt{\gamma} \left[ D \left( v^i - \frac{N^i}{N} \right) \right] \right\rbrace \right) = 0,
\end{equation}
momentum density:

\begin{multline*}\nonumber
\frac{1}{N \sqrt{\gamma}} \left( \frac{\partial}{\partial t} \left( \sqrt{\gamma} S_j \right) + \frac{\partial}{\partial x^i} \left\lbrace N \sqrt{\gamma} \left[ S_j \left( v^i - \frac{N^i}{N} \right) + p \delta_{j}^{i} \right] \right\rbrace \right)\\
= T^{0 0} \left( \frac{1}{2} N^k N^l \frac{\partial}{\partial x^j} \left( \gamma_{k l} \right) - N \frac{\partial}{\partial x^j} \left( N \right) \right) + T^{0 i} N^k \frac{\partial}{\partial x^j} \left( \gamma_{i k} \right) + \frac{S_k}{N} \frac{\partial}{\partial x^j} \left( N^k \right),
\end{multline*}
and relativistic energy density:

\begin{multline*}\nonumber
\frac{1}{N \sqrt{\gamma}} \left( \frac{\partial}{\partial t} \left( \sqrt{\gamma} \tau \right) + \frac{\partial}{\partial x^i} \left\lbrace N \sqrt{\gamma} \left[ \tau \left( v^i - \frac{N^i}{N} \right) + p v^i \right] \right\rbrace \right)\\
= T^{0 0} \left( N^i N^j K_{i j} + N^i \frac{\partial}{\partial x^i} \left( N \right) \right) + T^{0 i} \left( - \frac{\partial}{\partial x^i} \left( N \right) + 2 N^j K_{i j} \right) + T^{i j} K_{i j},
\end{multline*}
with primitive fluid variables ${\rho}$ (rest mass density), ${v^i}$ (three-velocity), and $p$ (pressure). The relativistic mass density $D$, three-momentum density ${S_i}$, and relativistic energy density ${\tau}$ are related to these primitive variables by:

\begin{equation*}
D = \rho W, \qquad S_i = \rho h W^2 v_i, \qquad \tau = \rho h W^2 - p - \rho W,
\end{equation*}
with $W$ being the Lorentz factor of the fluid and $h$ being its specific enthalpy:

\begin{equation*}
W = N u^0 = \frac{1}{\sqrt{1 - \gamma_{i j} v^i v^j}}, \qquad h = 1 + \frac{p}{\rho} \left( \frac{\Gamma}{\Gamma - 1} \right),
\end{equation*}
assuming an ideal gas equation of state with adiabatic index ${\Gamma}$. The three-velocity ${v^i}$ of the fluid is related to its four-velocity ${u^{\mu}}$ by:

\begin{equation*}
v^i = \frac{u^i}{N u^0} + \frac{N^i}{N}.
\end{equation*}
The right-hand-sides of these equations contain source terms involving the (perfect fluid) stress-energy tensor ${T^{\mu \nu}}$:

\begin{equation*}
T^{\mu \nu} = \rho h u^{\mu} u^{\nu} + p g^{\mu \nu},
\end{equation*}
and the extrinsic curvature tensor ${K_{i j}}$:

\begin{align*}
K_{i j} &= - \frac{1}{2 N} \left( \frac{\partial}{\partial t} \left( \gamma_{i j} \right) + \nabla_i N_j + \nabla_j N_i \right)\\
&= - \frac{1}{2 N} \left( \frac{\partial}{\partial t} \left( \gamma_{i j} \right) + \frac{\partial}{\partial x^i} \left( N_j \right) - \Gamma_{i j}^{k} N_k + \frac{\partial}{\partial x^j} \left( N_i \right) - \Gamma_{j i}^{k} N^k \right),
\end{align*}
where the particular form of these source terms assumes that the ADM Hamiltonian and momentum constraint equations are satisfied. The homogeneous part of this equation system is integrated in a locally-flat tetrad basis using a Roe-type finite volume solver, while the source terms are integrated using a four-stage, third-order, strong stability-preserving Runge-Kutta (SSP-RK) scheme\cite{gottlieb_strong_2001}. In all of the above, ${\gamma}$ represents the determinant of the spatial metric tensor ${\gamma_{i j}}$.

\subsection{Equations of General Relativistic Electromagnetism}
\label{sec:grem_equations}

\textsc{Gkeyll} solves the equations of general relativistic electromagnetism in hyperbolic, conservation law form, neglecting the elliptic constraint equations and the charge/current source terms (i.e. we take ${\rho = 0}$, ${\mathbf{J} = \mathbf{0}}$, since Maxwell's equations are not being coupled to any type of matter field), consisting of evolution equations for the electric and magnetic fields:

\begin{equation*}
- \frac{\partial \mathbf{D}}{\partial t} + \nabla \times \mathbf{H} = 0, \qquad \frac{\partial \mathbf{B}}{\partial t} + \nabla \times \mathbf{E} = 0,
\end{equation*}
where ${\mathbf{D}}$ and ${\mathbf{B}}$ represent specifically the electric and  magnetic fields \textit{as perceived by Eulerian observers}. In the above, ${\nabla}$ represents the covariant derivative operator on spacelike hypersurfaces, and ${\mathbf{E}}$ and ${\mathbf{H}}$ represent auxiliary fields defined by the vacuum constitutive relations:

\begin{equation*}
\mathbf{E} = N \mathbf{D} + \mathbf{N} \times \mathbf{B}, \qquad \mathbf{H} = N \mathbf{B} - \mathbf{N} \times \mathbf{D},
\end{equation*}
respectively. Once again, this equation system is integrated in a locally-flat tetrad basis using a Roe-type finite volume solver\cite{gorard2025}.

\end{appendices}

\end{document}